\documentclass[aps,prl,preprint,superscriptaddress,floatfix]{revtex4-1}
\usepackage{graphicx}
\usepackage{dcolumn}
\usepackage{bm}
\usepackage{amsmath,amsfonts}
\usepackage{color}
\usepackage{todonotes}
\usepackage{soul}
\usepackage{pdfpages}

\newcommand{\strikethrough}[1]{\st{#1}}

\begin{document}



\title{Edge phonons in black phosphorus}

\author{H. B. Ribeiro}
\affiliation{MackGraphe - Graphene and Nanomaterials Research Center, Mackenzie Presbyterian University, 01302-907 S\~ao Paulo, Brazil}

\author{C. E. P. Villegas}
\affiliation{Instituto de F\'isica Te\'orica, Universidade Estadual Paulista Julio de Mesquita Filho (UNESP), S\~ao Paulo SP, 01140-070, Brazil}

\author{D. A. Bahamon}
\affiliation{MackGraphe - Graphene and Nanomaterials Research Center, Mackenzie Presbyterian University, 01302-907 S\~ao Paulo, Brazil}

\author{D. Muraca} 
\affiliation{Instituto de F\'isica “Gleb Wataghin” (IFGW), Universidade Estadual de Campinas, Campinas, S\~ao Paulo, Brazil}

\author{A. H. Castro Neto}
\affiliation{Centre for Advanced 2D Materials and Graphene Research Centre, National University of Singapore, Singapore 117546, Singapore}

\author{E. A. T. de Souza}
\affiliation{MackGraphe - Graphene and Nanomaterials Research Center, Mackenzie Presbyterian University, 01302-907 S\~ao Paulo, Brazil}

\author{A. R. Rocha}
\affiliation{Instituto de F\'isica Te\'orica, Universidade Estadual Paulista Julio de Mesquita Filho (UNESP), S\~ao Paulo SP, 01140-070, Brazil}

\author{M. A. Pimenta}
\affiliation{Departamento de F\'isica, UFMG, 30123-970 Belo Horizonte, Brazil}

\author{C. J. S. de Matos}
\email{cjsdematos@mackenzie.br}
\affiliation{MackGraphe - Graphene and Nanomaterials Research Center, Mackenzie Presbyterian University, 01302-907 S\~ao Paulo, Brazil}


\date{\today}

\begin{abstract}
Exfoliated black phosphorus has recently emerged as a new two-dimensional crystal that, due to its peculiar and anisotropic crystalline and electronic band structures, may have potentially important applications in electronics, optoelectronics and photonics.
Despite the fact that the edges of layered crystals host a range of singular properties whose characterization and exploitation are of utmost importance for device development, the edges of black phosphorus remain poorly characterized. In this work, the atomic structure and the behavior of phonons near different black phosphorus edges are experimentally and theoretically studied using Raman spectroscopy and density functional theory calculations. Polarized Raman results show the appearance of new modes at the edges of the sample, and their spectra depend on the atomic structure of the edges (zigzag or armchair). Theoretical simulations confirm that the new modes are due to edge phonon states that are forbidden in the bulk, and originated from the lattice termination rearrangements. 
\end{abstract}


\maketitle


Black phosphorus (BP) is a thermodynamically stable allotrope of phosphorus that exhibits a layered structure, from which two-dimensional (2D) crystals can be obtained by means of mechanical exfoliation. Similar to graphene and MoS$_2$-like transition metal dichalcogenides (TMDs), the edges of BP can present armchair or zigzag atomic structures \cite{0957-4484-26-23-235707}. These different  types of crystallographic edges in graphene and TMDs are known to yield different magnetic, electronic and optical properties \cite{PhysRevB.73.205408,zhang2014,1_son_cohen_louie_2006,2_li_zhou_zhang_chen_2008,3_castro_peres_lopes}. Few studies, however, have reported on the edges of the lamellar puckered structure of BP \cite{morita1986semiconducting,sugai1985raman,cartz1979effect}, and its 2D counterpart, phosphorene \cite{koenig2014electric,li2014black,qiao2014high,10.1021/nl502892t,1_peng_copple_wei_2014,
0957-4484-26-23-235707,2015arXiv150603093P}. Previous theoretical simulations on phosphorene nanoribbons have demonstrated relaxation in the three atomic rows that are closest to the edges, and this rearrangement changes the mechanical properties of the material\cite{0957-4484-26-23-235707}. Density functional theory (DFT) calculations have also shown that
the nanoribbons can present a semiconductor or metallic character depending on the type and functionalization of the edges due to the appearance of electronic edge states\cite{1_peng_copple_wei_2014}. Such states have also been experimentally probed in single-layer phosphorene with scanning tunneling spectroscopy \cite{10.1021/nl502892t}. 

While little work has so far been done on phonon edge states, they are expected to have a significant impact on physical properties such as thermal conductivity and heat dissipation, which even removed from edges have been shown to be highly anisotropic \cite{0957-4484-26-5-055701,C4CP04858J,1_luo_maassen_deng,2_lee_yang_suh_yang_lee}. Using a Boltzmann thermal transport equation, Jain and McGaughey \cite{1_jain_mcgaughey_2015} were able to show that the optical phonon modes give a significant contribution (from 15 to 27\% depending on the direction of transport) to the thermal conductivity of phosphorene. Furthermore, Zhang et al. \cite{1_zhang_liu_cheng_wei} simulated the thermal transport properties of phosporene nanoribbons with different widths and different orientations and showed that the thermal conductivity is reduced at the edges. This kind of behavior is also observed on the edges of graphene nanoribbons, and other semiconductors such as silicon and germanium. It is therefore clear that a full characterization of edge effects is imperative, not only from the structural and fundamental physics point of view, but also for the technological applications of this novel material.

\begin{figure}
\centering
\includegraphics[width=0.9\textwidth]{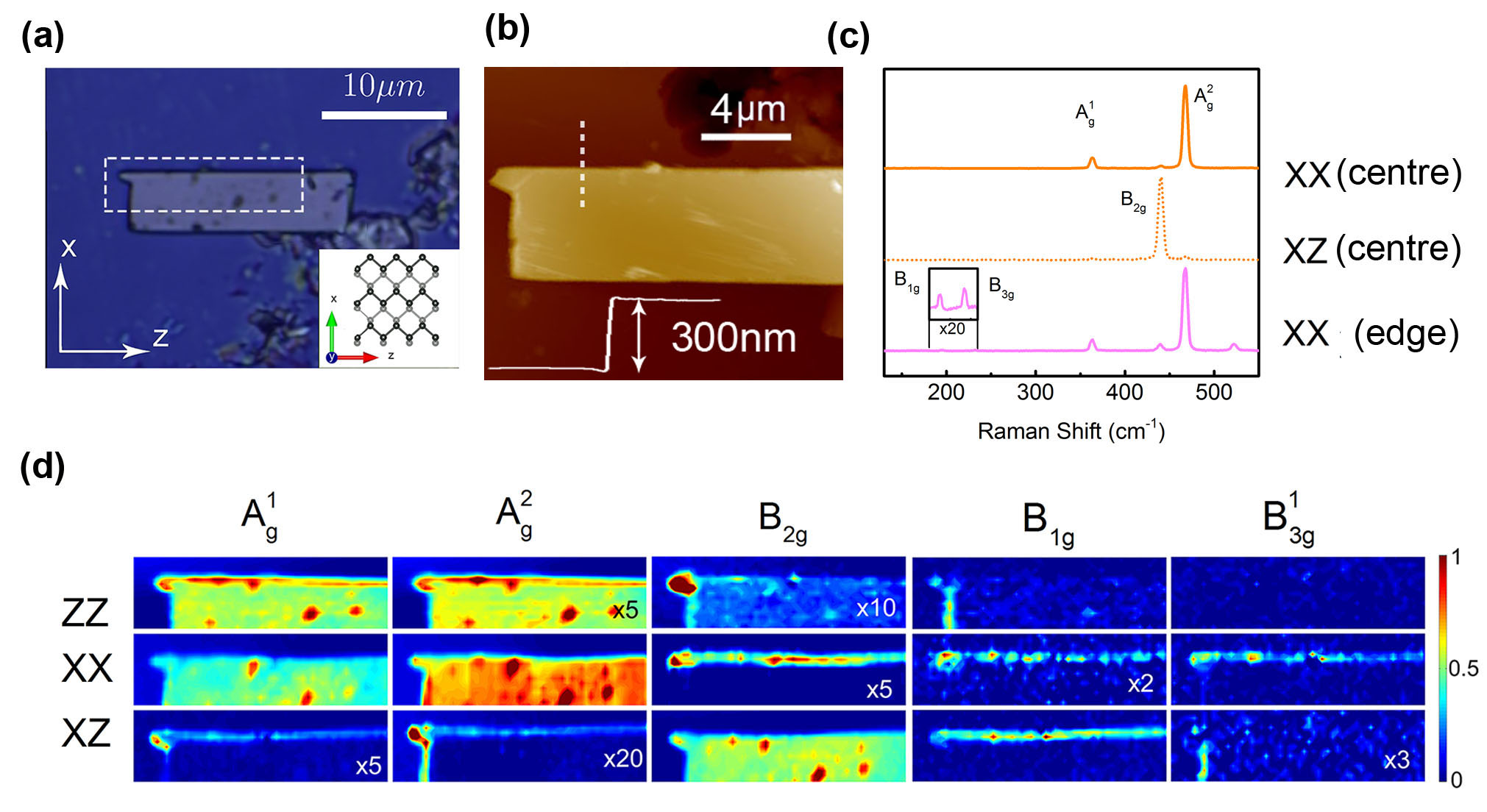}
\caption{\label{fig:fig1} \footnotesize\textbf{Optical image, Raman spectra and hyperspectral images of an exfoliated flake of black phosphorus with armchair and zigzag edges.} (a) Optical microscope image of the measured flake; the white dashed rectangle defines the analysed area. The corresponding crystal orientation and crystallographic x and z axes are presented in the inset. (b) AFM image of the measured flake (c) Polarized Raman spectra at the center of the flake with the XX (top) and XZ (middle) scattering configurations and at an armchair edge (bottom) with the XX configuration, where the first (second) index represents the polarization of the incident (scattered) light. (d) Hyperspectral Raman intensity images of all observed modes: A$_g^1$ , A$_g^2$ , B$_{2g}$ , B$_{2g}$ and B$_{3g}^1$ (columns), for the different scattering configurations indicated on the left side. Band intensities are represented by the color bar and were normalized to the highest value of each mode. When necessary, for better visualization, the intensity was multiplied by a factor, as indicated at the bottom-right corner of each image.}
\end{figure}

In this work we report on an experimental study of armchair and zigzag edges in exfoliated BP flakes by polarized Raman spectroscopy \cite{ribeiro215unusual}, conducted by hyperspectral imaging (see Figure \ref{fig:fig1}(a)). The results show the breakdown of the symmetry selection rules for the Raman active A$_g^1$, A$_g^2$ and B$_{2g}$ modes at the sample edges, and the presence in the spectra of the otherwise not allowed B$_{1g}$ and B$_{3g}^1$ modes. These unexpected results are found to be dependent on the edge type.
Our DFT calculations show that the experimental results can only be explained by the rearrangement of the atomic structure at the edges of the crystal \cite{0957-4484-26-23-235707}, leading to the appearance of edge phonon modes.
The numerically obtained Raman tensor elements near the distorted edges explain the appearance of forbidden symmetry modes in the polarized Raman spectra, and our results show that edge phonon states arise as a consequence of reconstructed atomic structure at the edges.

\section*{Results and Analysis}

Polarized Raman hyperspectral images were obtained with a confocal Raman spectrometer in a $300$-nm-thick rectangular flake of BP, shown in Figure \ref{fig:fig1}(a) and (b). The edges have well defined and known atomic structures, with the $z$ axis parallel to the zigzag edge and the $x$ axis along the armchair edge (see Methods section). The two uppermost spectra in Figure \ref{fig:fig1}(c) were recorded at the centre of the sample, using the XX and XZ scattering configurations, where the first (second) index corresponds to the polarisation of the incident (scattered) light. As expected by the symmetry selection rules\cite{loudon}, the totally symmetric A$_g^1$ and A$_g^2$ modes appear in the XX spectrum, centered at $360$ cm$^{-1}$ and $470$ cm$^{-1}$, respectively, while the B$_{2g}$ mode appears in the XZ spectrum, at $440$ cm$^{-1}$. The spectrum at the bottom of Figure \ref{fig:fig1}(c) was obtained at the armchair edge of the flake, using the XX scattering configuration. Note, in this case, that not only the B$_{2g}$ mode becomes visible, but we can also observe additional peaks at $190$ cm$^{-1}$ and $230$ cm$^{-1}$, which are identified as the B$_{1g}$ and B$_{3g}^1$ modes, according to frequency assignment from previous work \cite{sugai1985raman}. These two modes are not allowed by group theory analysis when the light polarisation lies in the $xz$ plane\cite{ribeiro215unusual,loudon}, and the reason for their observation will be discussed later in this work.

The complete study of the polarized Raman spectra of the analysed BP flake is shown through hyperspectral images in Figure \ref{fig:fig1}(d), where the Raman maps correspond to the spatially-resolved intensity of modes A$_g^1$, A$_g^2$, B$_{1g}$, B$_{2g}$ and B$_{3g}^1$ in the XX, ZZ and XZ scattering configurations (ZX maps are similar to those of the XZ configuration and are shown in the Supplementary Information). The existence of high-intensity localized dots in the hyperspectral images corresponds to a broad luminescence background, rather than Raman enhanced signals, possibly associated with defective or oxidized regions in the sample, seen as dark dots in Figure \ref{fig:fig1}(a). This is clear from the Raman spectrum taken at these dots shown in Fig. S2 of the Supplementary Information).

Let us first discuss the Raman images far from the sample edges. Notice in Fig. \ref{fig:fig1}(d) that all results for this case are consistent with group theory predictions: the totally symmetric A$_g^1$ and A$_g^2$ modes appear in the parallel polarisation configurations (ZZ and XX), the B$_{2g}$ is visible only in the crossed polarisation configuration (XZ), and the B$_{1g}$ and B$_{3g}$ modes are absent (they should only appear in the XY/YZ and YZ/ZY scattering configurations, respectively). 
By contrast, at the edges of the sample, we can observe several features that are in disagreement with \strikethrough{the} group theory-based symmetry analysis. Firstly, the totally symmetric  A$_g^1$ and A$_g^2$ modes appear in the crossed polarisation configuration, XZ.
More precisely, the A$_g^1$ mode signal is observed at the zigzag edge and a very weak signal is also seen at the armchair edge. For the A$_g^2$ mode, edge signals are observed for both the zigzag and armchair edges. The most significant anomalous signals in the XZ configuration are observed at the zigzag edge for the A$_g^1$ mode and at the armchair edge for the A$_g^2$ mode. Another anomalous result is observed for the B$_{2g}$ mode, which appears in the Raman images for the XX configuration at the zigzag edge and for the ZZ configuration at the armchair edge. 

Finally, Figure \ref{fig:fig1}(d) also shows the presence of the B$_{1g}$ or the B$_{3g}^1$ modes at the edges in all scattering configuration spectra (see two last columns). As already mentioned, the observation of these modes is not expected when both the incident and analysed radiations are polarized in the $xz$ plane, because the Raman tensors for these modes only exhibit $xy$ (for the B$_{1g}$ mode) and $yz$ (for the B$_{3g}^1$ mode) non-zero components. Indeed, as shown in Fig. \ref{fig:fig1}(d), they are not present away from the edges in the studied flake. The hyperspectral images in  Figure \ref{fig:fig1}(d) (and in the Supplementary Information) reveal that the B$_{1g}$ mode is clearly observed at the zigzag edge in both the ZX and XZ polarisation configurations, with a faint but clear signal also observed in the XX configuration. The armchair edge also exhibits a signal of this mode in the ZZ configuration.  The B$_{3g}^1$ mode, meanwhile, is observed at the armchair edge with both XZ and ZX configurations, and at the zigzag edge with a XX polarisation configuration.

Similar measurements were carried out in thinner flakes, with thicknesses of 30 nm and 6 nm, as determined by AFM; the same trends were observed and are presented in the Supplementary Information. The anomalous behavior for the totally symmetric A$_g$ modes and the $B_{2g}$ mode are observed at the edges of the sample, but the relative intensity of the edge phonons with respect to the allowed bulk phonons decreases with decreasing thickness. Accordingly, the $B_{1g}$ and $B_{3g}$ modes, which were already very weak in the 300-nm thick flake, could not be distinguished for the two thinner flakes.

All the anomalous results shown at the edges of the BP flake in Figure \ref{fig:fig1}(d) can be summarized considering the elements of the Raman tensors for each symmetry mode. For the totally symmetric A$_g$ modes, off-diagonal components appear in the Raman tensor and give rise to a signal when the analysed scattered light is orthogonal to the incident polarisation (XZ configuration). On the other hand, 
the Raman tensor of the B$_{2g}$ acquires diagonal components and, thus, a signal can be observed when the analysed scattered polarisation is parallel to the incident polarisation (ZZ and XX configurations). Finally, the B$_{1g}$ and B$_{3g}$ Raman tensors, which usually have only $xy$ and $yz$ elements, respectively, acquire $xx$, $zz$ and $xz$ components at the edges of BP, leading to the observed anomalous signals. As shown in the following, the changes in the Raman tensors result from the previously reported lattice reconstructions near the edges\cite{10.1021/nl502892t,0957-4484-26-23-235707}, giving rise to edge phonons with different symmetries.

\begin{figure}
\centering
\includegraphics[width=0.9\textwidth]{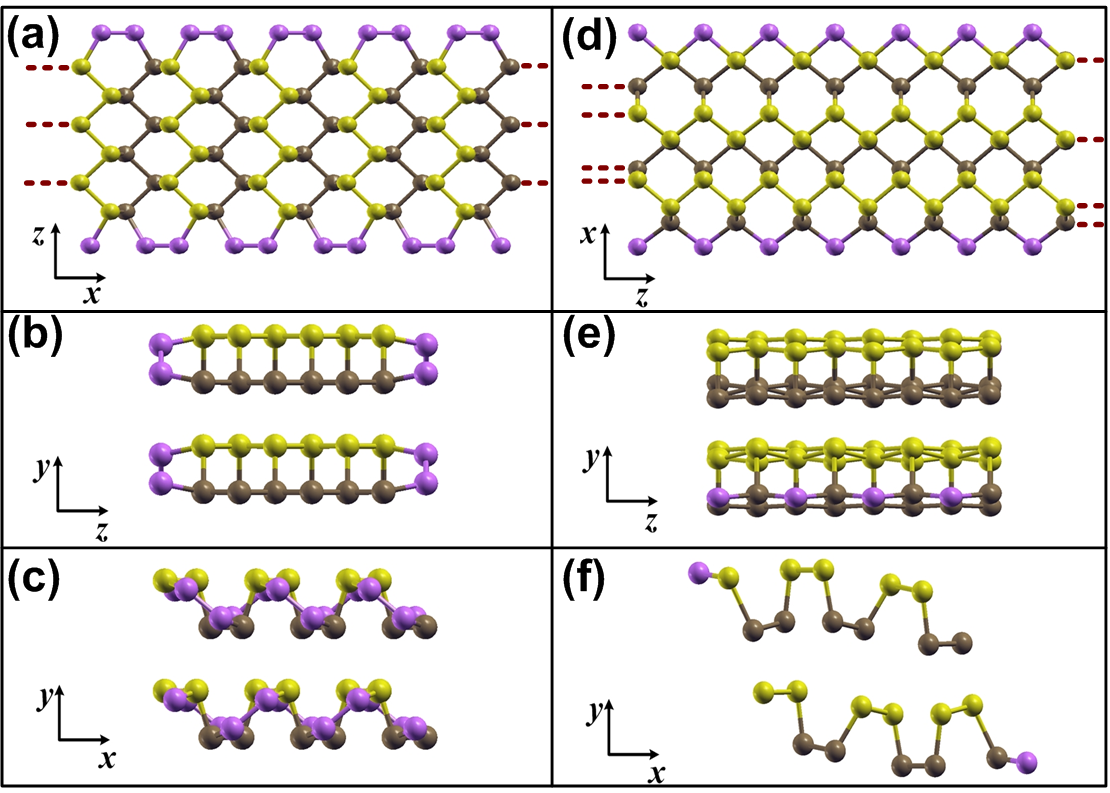}
\caption{\label{fig:fig5}\footnotesize\textbf{Black phosphorus slabs studied by DFT calculations and their corresponding relaxed edges.} (a) One layer of a BP slab with armchair edges, and the corresponding projections on the (b) $yz$ plane, and (c) $xy$ plane of the unit cell used. (d)  One layer a BP slab with zigzag edges, and corresponding projections on the (e) $yz$ plane, and (f) $xy$ plane.
The atoms at the edges of each slab are shown in purple to highlight the atomic rearrangements. The structures are repeated periodically in the $y$ direction in both cases, as well as in the $x$ direction for the armchair slab, and in the $z$ direction for the zigzag case.}
\label{fig:fig2}
\end{figure}

\begin{figure}[ht]
\centering
\includegraphics[width=0.9\textwidth]{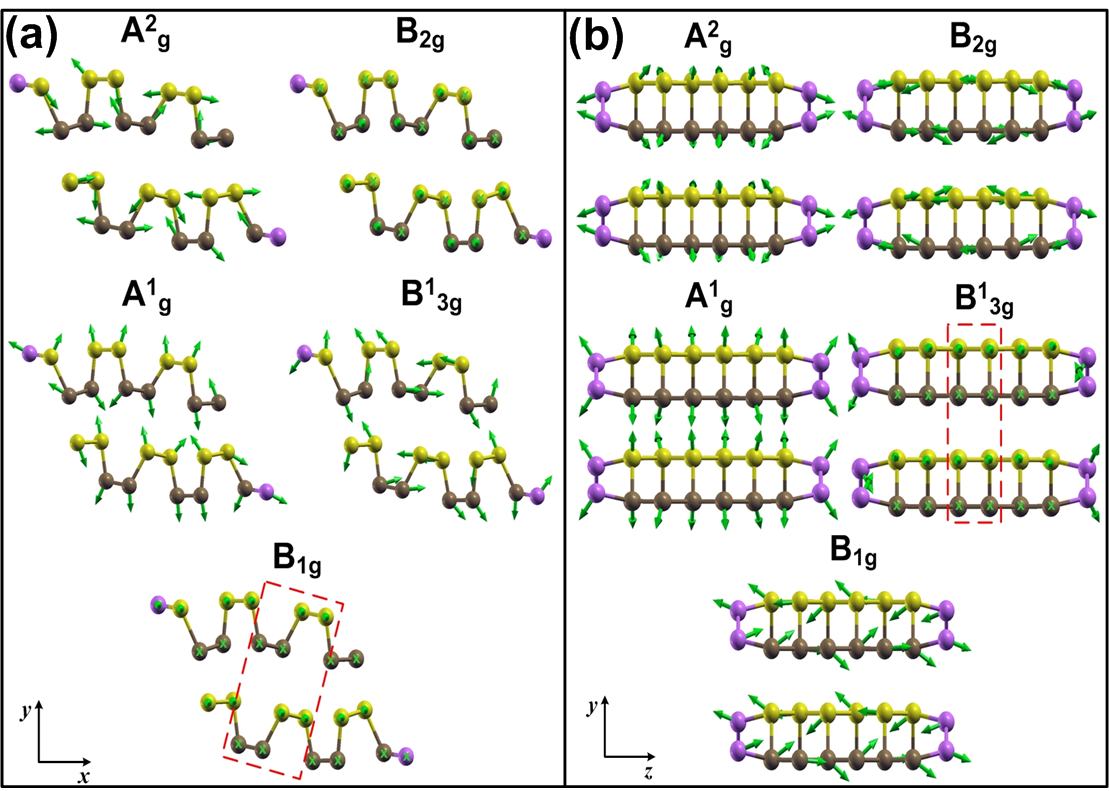}
\caption{\label{fig:fig6}\footnotesize\textbf{Atomic displacements corresponding to the observed Raman active modes at the edges.} (a) zigzag edge; (b) armchair edge. The atoms at the edges of each slab are shown in purple.}
\label{fig:fig3}
\end{figure}

\begin{figure}[ht]
\includegraphics[width=\textwidth]{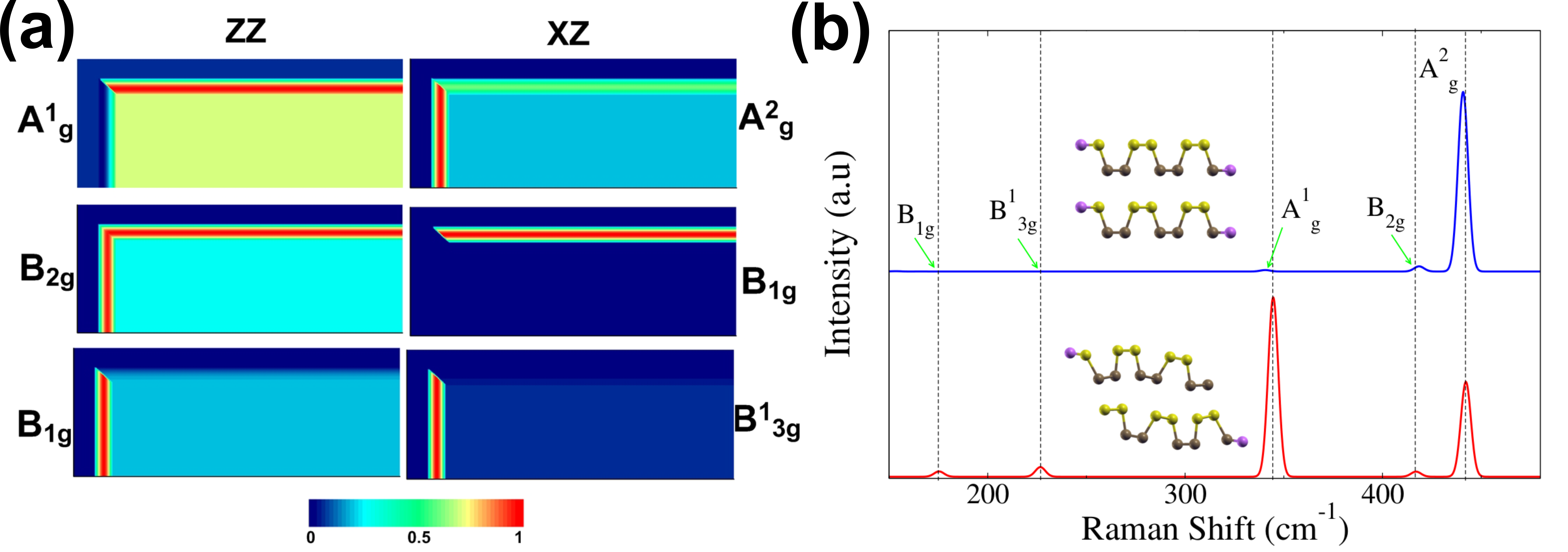}
\caption{\footnotesize\textbf{Simulated Raman images and spectra.} (a) Representative simulated Raman hyperspectral images for selected modes and scattering configurations. In each image, the top and left-hand-side regions correspond to the silicon substrate. (b) Unpolarized Raman spectra for an AB-stacked zigzag-terminated slab, with (red) and without (blue) edge reconstruction (structure relaxation).}\label{fig4}
\end{figure}

In order to elucidate the physical origin of the anomalies in the polarised Raman spectra at the edges of black phosphorus, we performed {\it ab initio} density functional theory calculations to determine the rearrangement of the atomic structure at the crystal terminations and its effect on the phonon modes, for both zigzag and armchair edges. Figure \ref{fig:fig2} shows different perspectives of the relaxed unit cell used in this work. Since previous experimental studies performed with samples of the same origin\cite{1_castellanos-gomez} show that the BP crystals are AB stacked, we have considered infinitely thick and $\sim 13$~\AA~ wide BP slabs consisting of AB-stacked layers \cite{C4CS00257A,sengupta2015effect,physRevB.92.165406}, with lateral surfaces terminated by either zigzag or armchair edges. 

Starting with the armchair termination (Figs. \ref{fig:fig2}(a-c)), we note a clear reconstruction of the edge. Figure \ref{fig:fig2}(a) shows a top view of one of the layers of BP after relaxation. Here one can note that atoms along the edges have significantly different positions compared to their counterparts inside the slab. The lateral view in Figures  \ref{fig:fig2}(b-c) clearly show the reconstruction, where the edge atoms within the same layer move closer together. Restructuring is more subtle in the case of the zigzag edges (Figs. \ref{fig:fig2}(d-f)). The top view of Figure \ref{fig:fig2}(d) shows one layer of the zigzag-terminated slab. From this perspective the most noticeable feature is the stretching and contraction of the puckered structure along the $x$ direction, {\it i.e.}, one can notice that the distance between the projected atoms in the hexagons clearly varies depending on the region of the slab considered. This effect is also clear in the lateral view of Figure \ref{fig:fig2}(e), where only small deviations from the bulk layered material are observed (see Supplementary Information). The projection on the $xy$ plane represented in Figures \ref{fig:fig2}(c) and \ref{fig:fig2}(f) shows a shear dislocation of the layers forming the unit cell, as well as a distortion of the edge atomic structure.

We calculated all the normal phonon modes from each of the relaxed structures. Figures \ref{fig:fig3}(a) and (b) show the atomic displacement vectors corresponding to the A$_g^1$, A$_g^2$, B$_{2g}$, B$_{1g}$ and B$_{3g}^1$ modes for slabs with zigzag and armchair surfaces, respectively (see Supplementary Information). These displacements can be compared to those for the bulk material (see Supplementary Information). These modes were assigned by comparing their frequencies and displacements with those of bulk BP. One can note that the atomic displacements in the middle of the slab somewhat resemble those observed for the bulk modes. Furthermore, they also correspond to the modes seen in the interior of the sample in the experiments. On the other hand, near the surface, edge reconstruction causes dramatic changes on the atomic displacement vectors. This result is independent of the slab width, and is in fact the physical mechanism behind the anomalous behaviour of the phonon modes in the polarised Raman spectra.
In particular, we clearly observe, for the B$_{3g}^1$ mode of the armchair slab [Figure \ref{fig:fig3}(b)], that atomic displacements in the interior of the slab take place on the $xz$ plane whereas the edge atoms present out-of-plane vibrations.

In order to compare the experimental Raman results with the theoretically predicted atomic vibrations at the BP edges, we used the PHonon code integrated into the Quantum Espresso package to obtain the Raman tensors within the Placzek approximation. All Raman intensities for each edge - considering different polarisations - are presented in the Supplementary Information where they are analysed in details. The Raman intensity of a specific edge phonon mode for a given scattering configuration can be traced back to the elements of its Raman tensor. We observed for the relaxed structures that new elements appear in the Raman tensor due to the sum over the atomic displacements of the normal vibrational modes at the edges with different atomic structures. These new elements are responsible for the anomalous Raman peaks at the edges, edge phonon states.

By analysis of the simulated Raman tensors in bulk BP crystals and BP slabs, theoretical hyperspectral images could be obtained, corresponding to a flake with similar edge terminations to the experimental one, and results are shown in Figure \ref{fig4}(a). In each image, the top and left-hand-side regions correspond to the silicon substrate, as in the experimental images shown in Figure \ref{fig:fig1}. There is a noticeable resemblance between the calculated and the experimental maps in Figure \ref{fig:fig1}(d). First, we could observe the appearance of the A$_g$ and B$_{2g}$ modes in the XZ and ZZ spectra, respectively. Moreover, the activation of the forbidden modes B$_{1g}$, for the zigzag surface, and $B_{3g}^1$, for the armchair surface, with the  XZ scattering configuration reveals that this is not a simple effect induced by confinement. Finally, we can conclude that the relaxed zigzag and armchair surfaces effectively describe the physical phenomena experimentally observed at the edges. 

In order to further assess the mechanism behind the anomalous appearance or mode enhancement, Figure \ref{fig4}(b) shows a comparison between the Raman spectra of the perfectly defined zigzag surface and the Raman spectra of the relaxed structure, where the atomic structure is reconstructed at the edge. Firstly, we can observe that the $B_{1g}$ and $B_{3g}^1$ modes appear only in the spectrum of the relaxed structure.  Furthermore, only in the relaxed structure a significant increase in the $A_g^1$ mode intensity occurs. In general, due to broken symmetry at the edges, forbidden scattering becomes allowed and new phonon modes can appear at the edges, as in the case of D band in graphene edges \cite{PhysRevLett.93.247401}. However, as shown in Figure \ref{fig4}(b), this is not the case for BP. The observed edge phonon modes arise as a consequence of the reconstruction of the atomic structure near the edges, and not simply due to the breakdown of the translational symmetry.

\section*{Conclusion}

We have used polarised Raman hyperspectral imaging to study flakes of exfoliated black phosphorus with thicknesses spanning from 300 nm down to 6 nm, and exhibiting well defined zigzag and armchair atomic edge structures. The polarisation of the incident and scattered light were changed with respect to the edge directions in order to distinguish the bulk phonon modes of different symmetries. Whereas the spectra obtained in the interior of the sample could be well described by group theory analysis based on the Raman tensors for each symmetry mode, anomalous results were observed at both the zigzag and armchair edges. The totally symmetric A$_g$ modes were observed in the crossed polarisation configurations, and the B$_{2g}$ symmetry mode was observed in the parallel polarisation configurations, whereas the B$_{1g}$ and B$_{3g}^1$ symmetry modes, not observed in the sample interior, were observed near the edges. These observations clearly indicate the presence of edge phonon modes in black phosphorus. The experimental results were explained by density functional theory (DFT) calculations that predicted the local rearrangement of atoms near the edges, as well as changes in the atom displacements corresponding to each Raman mode, leading to the emergence of additional Raman tensor elements.
The resulting simulated hyperspectral Raman images of the studied flake were shown to be in excellent agreement with the experimental data and, therefore, capture the existence of edge phonons in BP. This work, thus, improves the understanding of the phonon behaviour and of the morphology at the edges of black phosphorus, and will contribute to the development of optimized photonic, phononic and electronic devices using this material, especially those employing edges and nanoribbon-like geometries.

\section{methods}

\subsection{Experimental Methods}

Raman images were obtained in a confocal Raman spectrometer (WITec Alpha 300R) using a 488 nm (2.54 eV) laser line and a 100$\times$ objective lens. To avoid laser induced thermal effects the optical power was kept constant at 1 mW during all measurements. Polarised Raman spectra were obtained by inserting an analyser, aligned either parallel or perpendicular to the polarisation of the incident light, before the spectrometer entrance. A multimode fiber, positioned after the analyser, was able to depolarise the light sent to the spectrometer, effectively removing the polarisation dependence of the diffraction grating. 

Flakes of BP with a rectangular geometry were produced by mechanical exfoliation onto a Si substrate with a 300-nm-thick SiO$_2$ cover layer. The crystal orientation and the edge character (zigzag or armchair) were obtained from linear dichroism, HRTEM and electron diffraction measurements (see Supplementary Information). The thicknesses of the studied flakes were 300, 30 and 6 nm, as measured by atomic force microscopy. The flake shown in Figure \ref{fig:fig1} exhibits a $300$-nm thickness. In order to avoid oxidation or other environmental degradation effects, all measurements were carried out with a continuous flow of nitrogen gas.

During experiments the incident light remained polarised along the horizontal direction of Figure \ref{fig:fig1} and the sample was rotated $90$ degrees so that the polarisation could be parallel or perpendicular to each edge. However, for simplicity, all data are shown in the sample's reference frame, within which the incident light is polarised along the $x$ or $z$ axes.

\subsection{Theoretical Methods}

In all calculations, plane-wave density functional theory \cite{dft1,dft2} was employed to obtain the electronic ground-state using the Perdew-Burke-Ernzerhof (PBE)\cite{pbe} generalized gradient exchange-correlation functional,
currently implemented in the Quantum-Espresso package\cite{QE}. Van der Waals corrections
within the semi-empirical dispersion scheme (PBE-D) proposed by Grimme\cite{grimme} were also included. Norm-conserving pseudopotentials with 3s3p states were adopted to describe electronic states of phosphorus. The Brillouin zone was mapped within the Monkhorst-Pack scheme 
using 7 $\times$ 7 $\times$ 7 k-sampling grid for the bulk and a 10 $\times$ 7 $\times$ 1 (1 $\times$ 8 $\times$ 8) grid for the 
armchair (zigzag) slab samples. The kinetic energy cutoff was set at 90 Ry and 100 Ry for bulk and slab geometries, respectively. Furthermore a vacuum region of 16 \AA \ was adopted for the supercell related to the surfaces. The structures were fully optimized to their equilibrium positions with forces smaller than 0.002 eV/\AA \ and the unit cells were relaxed to a target pressure of 0.2 kbar.

The armchair structure was constructed by cutting the bulk black phosphorus along two $xy$ planes. Thus, the slab is composed by 8 atomic rows that extend along the $z$-axis, yielding a width of $\sim$13 \AA. 
The zigzag edge was built by cutting the black phosphorus crystal along two $yz$ planes. The width of the slab is $\sim$13.5 \AA \ and the system consists of 12 atomic rows per layer that are parallel to the $y$-direction. In all cases, in order to obtain an AB-stacked slab while maintaining an orthorhombic unit cell, we considered a periodically repeated double layer of BP in our calculations.

Within density functional perturbation theory (DFPT) \cite{dfpt_ph,dfpt_rev,imple-raman,Lazzeri:2003yu}, the linear response approach can be used to calculate the dielectric tensor, the different vibrational frequencies, and corresponding normal modes.
Within the Placzek approximation\cite{Lazzeri:2003yu}, the Raman intensity for a particular normal mode
$\nu$ with frequency $\omega$, related to the nonresonant Stokes process, is obtained as
\begin{equation}
 I^{\nu} \propto |\textbf{e$_{i}$\ \ .} \ \  A^{\nu}\ \ \textbf{.}  \ \ \textbf{e$_{s}$}|^{2}\frac{1}{\omega_{\nu}}(n_{\nu}+1)
\end{equation}
where $\textbf{e$_{i}$} \ \ (\textbf{e$_{s}$}$) is the incident (scattered) light polarisation, $n_{\nu}$ the Bose-Einstein distribution, and $A^{\nu}$ is the Raman tensor with matrix elements
\begin{equation}\label{raman_tensor}
 A_{lm}^{\nu}=\sum\limits_{k \gamma} \frac{\partial^3 \mathcal{E}^{el}}{\partial E_{l} \partial E_{m} \partial u_{k \gamma}}\frac{w_{k \gamma}^{\nu}}{\sqrt{M_{k}}} ~,
\end{equation}
where $\mathcal{E}^{el}$ is the electronic energy of the system in the presence of a uniform electric field $E_{l(m)}$ along direction $l (m)$, $u_{k\gamma}$ corresponds to the $\gamma-th$ component of the normal mode on atom $k$, and $M_k$ is the corresponding atomic mass.
 
Once the Raman tensor is calculated, we are able to choose, according to the experimental setup, the direction of the incident (scattered) light and thus obtain the Raman intensities using equation \ref{raman_tensor}. 
The zigzag and armchair BP surfaces exhibit a larger number of the vibrational modes; the zigzag surface presents 72 vibrational modes (24 atoms in the unit cell) including active and inactive Raman modes, as well as confined modes. In order to identity the modes we first focused on a frequency interval around the frequency of the active bulk vibrational mode, and then compared the atomic displacement vectors with those found for the bulk modes. The Raman tensor for all the structures were calculated using the PHonon code\cite{QE}, currently integrated into the Quantum-Espresso package. In order to obtain the simulated hyperspectral image we considered, for the interior of the flake, the bulk Raman tensor and evaluated the elements corresponding to each scattering configuration. For the edges, we considered the values obtained for the Raman tensor elements of the slabs, with the corresponding edge termination and polarisation configuration. 

\section{Acknowledgements}

\begin{acknowledgments} This work is supported by FAPESP (SPEC project 2012/50259-8) and partially supported by the Brazilian Nanocarbon Institute of Science and Technology (INCT/Nanocarbono), Fapemig, CNPq, and MackPesquisa. H.B.R. acknowledges a CNPq scholarship, and C.E.P.V. and D.M. acknowledge FAPESP fellowships (2012/24227-1 and 2011/01235-6). A. R. R. acknowledges support from ICTP-SAIRF (FAPESP project 2011/11973-4) and the ICTP-Simons Foundation Associate Scheme. Computational support was provided by Grid-Unesp and CENAPAD-SP. A.H.C.N. acknowledges the National Research Foundation, Prime Minister Office, Singapore, under its Medium Sized Centre Programme and CRP award Novel 2D materials with tailored properties: beyond graphene (R-144-000-295-281). We thank C2NANO-Brazilian Nanotechnology National Laboratory (LNNano) at Centro Nacional de Pesquisa em Energia e Materiais (CNPEM)/MCT (project nos. 14825 and 14827) and Prof. Marcelo Knobel for the use of the TEM facilities.
\end{acknowledgments}

\bibliographystyle{naturemag}
\bibliography{bibnew3} %

\end{document}